\documentclass{aa}  
\usepackage{graphicx}
\usepackage{txfonts}

\def\gsimeq
{\hbox{\raise0.5ex\hbox{$>\lower1.06ex\hbox{$\kern-1.07em{\sim}$}$}}}
\def\lsimeq
{\hbox{\raise0.5ex\hbox{$<\lower1.06ex\hbox{$\kern-1.07em{\sim}$}$}}}

\begin{document}

   \title{Correlated modulation between the redshifted Fe~K$\alpha$ line \\ 
     and the continuum emission in NGC~3783}

   \subtitle{}

   \author{F. Tombesi
\inst{1,2}
\and
B. De Marco
\inst{3}
\and
K. Iwasawa
\inst{4}
\and
M. Cappi
\inst{1}
\and
M. Dadina
\inst{1}
\and
G. Ponti
\inst{1,2}
\and
G. Miniutti
\inst{5}
\and
G.G.C. Palumbo
\inst{2}
          }

   \offprints{F. Tombesi \\ \email{tombesi@iasfbo.inaf.it}}

   \institute{INAF-IASF Bologna, Via Gobetti 101, I-40129 Bologna, Italy 
\and
Dipartimento di Astronomia, Universit\`a degli Studi di Bologna, Via Ranzani 1, I-40127 Bologna, Italy 
\and
International School for Advanced Studies (SISSA), Via Beirut 2-4, I-34014 Trieste, Italy
\and
Max-Planck-Institut fur Extraterrestrische Physik, Giessenbachstrasse, D-85748 Garching, Germany 
\and
Institute of Astronomy, Madingley Road, Cambridge CB3 OHA, United Kingdom
}

   \date{Received 8 November 2006 / Accepted 20 March 2007}

 
  \abstract
   {}
{It has been suggested that X-ray observations of rapidly variable Seyfert 
galaxies may hold the key to probe the gas
orbital motions in the innermost regions of accretion discs around black 
holes and, thus, trace 
flow patterns under the effect of the hole strong gravitational field.}
   {We explore this possibility by re-analyzing the multiple 
\emph{XMM-Newton} observations of the Seyfert 1 galaxy NGC~3783.
A detailed time-resolved spectral analysis is performed down to 
the shortest possible time-scales (few ks) using ``excess maps" 
and cross-correlating light curves in different energy bands.}
   {In addition to a constant core of the Fe~K$\alpha$ line, we detected a 
variable and redshifted Fe~K$\alpha$ emission feature between
5.3--6.1 keV. The line exhibits a modulation on a time-scale of 
$\sim$27 ks that is similar to and in phase with a modulation of the
0.3--10 keV source continuum. The two components show a good correlation.}
   {The time-scale of the correlated variability of the redshifted Fe line 
 and continuum agrees with the local dynamical time-scale of 
the accretion disc at $\sim$10~$\mathrm{r_g}$ around a black hole with 
the optical reverberation mass $\sim$10$^7 \mathrm{M_{\odot}}$. Given the 
shape of the redshifted line emission and the overall X-ray variability 
pattern, the line is likely to arise from the relativistic region 
near the black hole, although the source of the few cycles of coherent 
variation remains unclear.}


   \maketitle
%

\section{Introduction}

The fluorescent iron K (Fe~K$\alpha$) emission line is considered
to be a useful probe of the accretion flow around the central black hole of an
active galaxy. In particular, due to the high orbital velocity and strong
gravitational field in the innermost regions of an accretion disc, its profile
shall be deformed by the concurrence of Doppler and relativistic shifts. The
resulting line is therefore broadened, with a red-wing extending towards
lower energies (e.g. Fabian et al. \cite{fabian}; Reynolds \& Nowak 
\cite{reynolds}). 
Detailed modelling of time-averaged spectra
have been used to obtain important estimates of the disc
ionization state, its covering factor and, at least for the
brightest and best cases, its emissivity law, inner radius
and BH spin (Brenneman \& Reynolds \cite{brenneman}; Miniutti et
al. \cite{miniutti2006}; Guainazzi et al. \cite{guainazzi}; 
Nandra et al. \cite{nandra2006}). It is also well 
established that time-resolved spectral analysis
is a fundamental tool if we want to understand not only
the geometry and kinematics of the inner accretion flow
but also its dynamics. Early attempts (i.e. Iwasawa et al. \cite{iwasawa1996}; 
Vaughan \& Edelson \cite{vaughan}; Ponti et al. \cite{ponti}; Miniutti et al. 
\cite{miniutti2004}) have clearly shown that the redshifted
component of the Fe~K$\alpha$ line is indeed
variable and that complex geometrical and relativistic effects
should be taken into account (Miniutti et al. \cite{miniutti2004}).\\
More recently, results of iron line variability have been
reported (Iwasawa, Miniutti \& Fabian \cite{iwasawa2004}; Turner et al.
\cite{turner}; Miller et al. \cite{miller2006}). These are consistent with 
theoretical studies on the
dynamical behaviour of the iron emission arising from localized
hot spots on the surface of an accretion disc (e.g. Dov{\v c}iak et al. 
\cite{dovciak}). Iwasawa, Miniutti \&  Fabian (\cite{iwasawa2004}), 
for example, measured a $\sim$25
ks modulation in the redshifted Fe~K$\alpha$ line flux in NGC~3516, 
which suggests that the emitting region is very close to the central black
hole. However, these are likely to be transient phenomena, since such spots
are not expected to survive more than a few orbital revolutions.
For this reason, it is inherently difficult to establish the observational
robustness of these type of models, if not by accumulating further
observational data.

   \begin{figure*}
   \centering
    \includegraphics[height=3cm,width=8cm,angle=270]{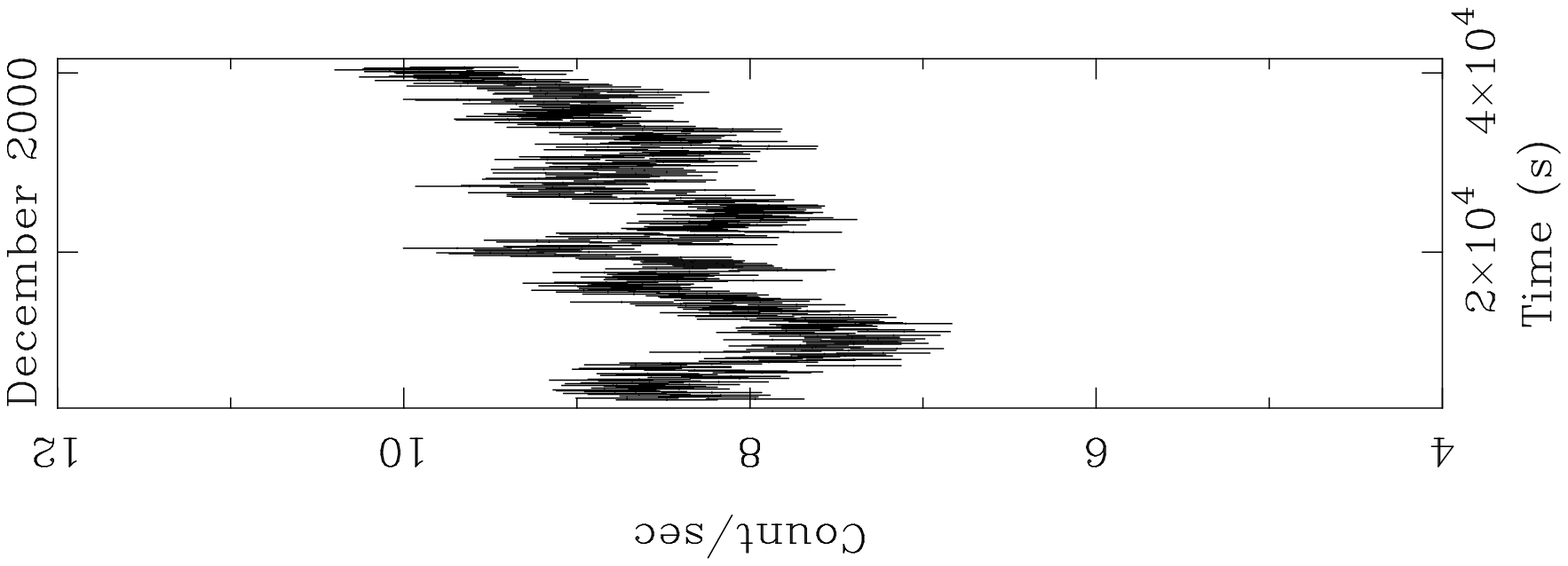}
    \includegraphics[height=14cm,width=8cm,angle=270]{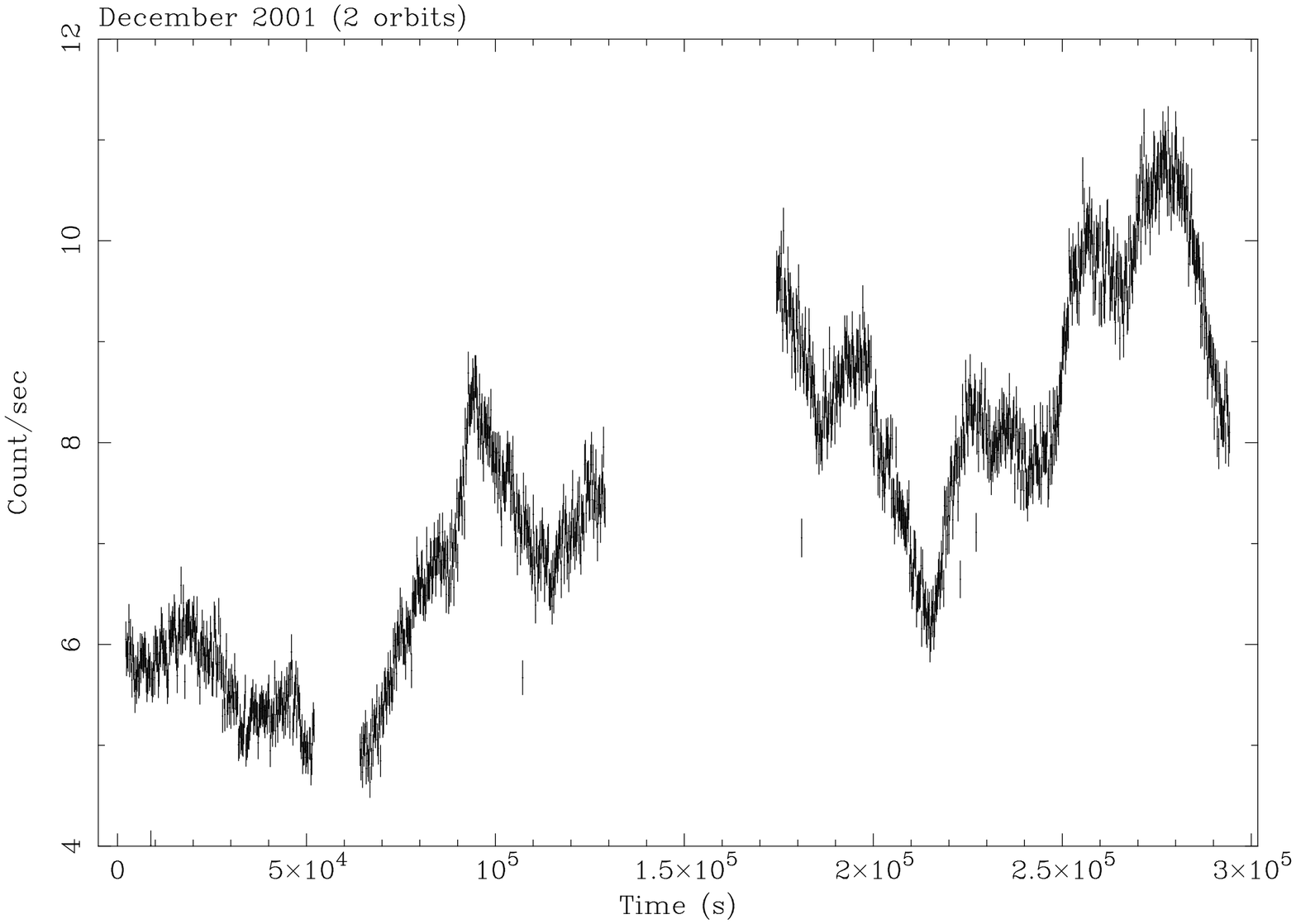}
   \caption{The X-ray light curves of NGC~3783 in the 0.3--10 keV 
band. 
\emph{Left panel}: light curve of the 2000 observation. 
\emph{Right panel}: light curves of the 2001a and 2001b observations.}
              \label{f1}
    \end{figure*}

Here we present results on the iron line variability in
the bright Seyfert galaxy NGC~3783 (z$\simeq$0.01) based on
\emph{XMM-Newton} data. This object has been taken as an example
in which multiple warm absorbers can mimic the
broad iron line feature (Reeves et al. \cite{reeves}), contrary to
the initial claim of the presence of a broad iron line emission
using ASCA data (Nandra et al. \cite{nandra1997}). However, the
recent study by De Marco et al. (\cite{demarco}) found evidence for
a transient excess feature in the \mbox{5--6} keV energy band,
interpreted as a redshifted component of the Fe~K line.
This result is also supported by a variability study by O'Neill \& Nandra 
(\cite{oneill}), who examined rms variability spectra of
a sample of bright active galaxies observed with \emph{XMM-Newton}. 
Given the above considerations, we re-examined
all the \emph{XMM-Newton} observations of NGC~3783 to perform
a comprehensive study of the iron line temporal evolution,
on the shortest possible time-scale.

\section{\emph{XMM-Newton} observations}

\emph{XMM-Newton} observed NGC~3783 on 2000 December 28--29 and on 
2001 December 17--21. The first observation (ID
0112210101) has a duration of $\sim$40 ks while the second
(ID 0112210201 and ID 0112210501, hereafter observation
2001a and 2001b respectively) lasts over two complete orbits
for a total duration of $\sim$270 ks. Only the EPIC pn
data are used in the following analysis because of the high
sensitivity in the Fe~K band. The EPIC pn camera was
operated in the ``Small Window" mode with the Medium
filter both during the 2000 and the 2001 observations. The
live time fraction is thus 0.7. The data were reduced using
the XMM-SAS v. 6.5.0 software while the analysis was
carried on using the \emph{lheasoft} v. 5.0 package. High background 
time intervals were excluded from the analysis.
The useful exposure time intervals are listed in Tab.~1,
together with the mean 0.3--10 keV count rate for each
observation. Only single and double events were selected.
Source photons were collected from a circular region of 56
arcsec radius, while the background data were extracted
from rectangular, nearly source-free regions on the detector.
The background is assumed to be constant throughout
the useful exposure. The 0.3--10 keV light curves are
shown in Fig.~\ref{f1} for each observation.

\section{Data analysis}

\subsection{Spectral features of interest and selection of the energy resolution}

\begin{figure}
   \centering
    \includegraphics[width=5cm,height=7cm,angle=270]{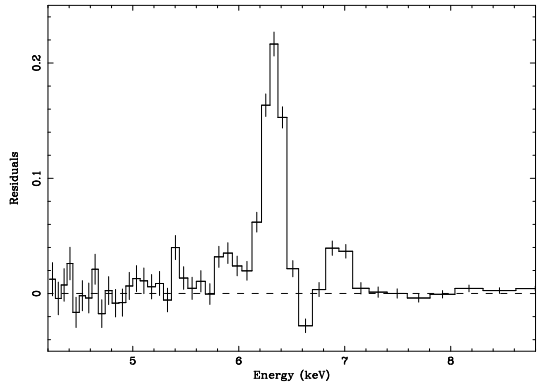}
   \caption{The 4--9 keV residuals against a simple power-law plus cold
 absorption continuum model for the spectrum of NGC~3783 during the 2001b 
observation. The data are obtained from EPIC pn.}
              \label{f2}
    \end{figure}

The time-averaged spectrum was analyzed using the
XSPEC v. 11.2 software package. For simplicity, we limited
the analysis to the 4--9 keV band. In this energy band,
we checked that the complex and highly ionized warm absorber
(with log$\xi$ and $N_{\rm H}$ up to $\sim$2.9 erg cm s$^{-1}$ and 
5$\times$10$^{22}$ cm$^{-2}$, Reeves et al. \cite{reeves}) shall not affect 
our conclusions below. The residuals against a simple power-law plus cold 
absorption continuum model for the 2001b observation, the longest continuous 
dataset available, are shown in Fig.~\ref{f2}. In this fit we 
excluded the Fe~K energy band (i.e. 5--7 keV) and the best fit parameters are 
\mbox{$(2.5\pm0.6)\times 10^{22}$ cm$^{-2}$} and $1.81\pm0.04$, for the 
absorber column density and power-law slope respectively.  

\begin{table}
\caption{Date, duration, useful exposure and mean EPIC pn
0.3--10 keV count rate for each \emph{XMM-Newton} observation of
NGC~3783.}
\label{tab:1}     
\centering          
\begin{tabular}{c c c c c}
\hline\hline             
 Obs. ID & Date & Duration & Exposure & $\langle \mathrm{CR} \rangle$ \\
 &       &  (ks) &  (ks) & (c/s)\\
\hline                       
 0112210101 & 2000 Dec 28--29 & 40.412  & 35 & 8.5 \\
 0112210201 & 2001 Dec 17--19 & 137.818 & 115 & 6.5 \\
 0112210501 & 2001 Dec 19--21 & 137.815 & 120 & 8.5 \\
\hline                        
\end{tabular}
\end{table}

Identified are four excess emission features: the main
Fe~K$\alpha$ core at $\sim$6.4 keV, a wing to the line core at around
6 keV, a peak at $\sim$7 keV (possibly Fe~K$\beta$) and a narrow
peak at $\sim$5.4 keV. Moreover we identified two absorption
features at $\sim$6.7 keV and $\sim$7.6 keV. When fitted
with Gaussian emission and absorption lines, all these features
are significant at more than $\sim$99\% confidence level.
Similar results were also obtained by a detailed analysis
with more complex models (Reeves et al. \cite{reeves}). We will
focus here on the analysis of the features variability properties.
The application of the excess map technique to the
identified absorption features did not give significant results,
thus, in the following, we will focus on the analysis
of emission features variability only. The 2001a observation
has been divided into two parts because of the gap in
the data between t$\sim$5$\times$10$^{4}$ s and t$\sim$6$\times$10$^{4}$
s. Since all the selected spectral features are comparable to the
CCD spectral resolution, we chose 100 eV for the energy
resolution of the excess maps.

\subsection{Selection of the time resolution}

In choosing the time resolution for the excess maps we
looked for the best trade-off between getting a sufficiently
short time-scale, in order to oversample variability, and
keeping enough counts in each energy resolution bin. We
first considered the 2001b observation, having the longest
and continuous exposure. Spectra were extracted during
different time intervals (1 ks, 2.5 ks and 5 ks) around the
local minimum flux state at t$\sim$215 ks. The required condition
is that each 100 eV energy bin in the 4--9 keV band
has to contain at least 50 counts. At the time resolution
of 2.5 ks we got $\sim$90 counts per energy bin at the energies
of the ``red" feature (5.3--5.4 keV), and $\sim$80 counts
per energy bin in the ``wing" feature energy band (5.8--6.1
keV).
Moreover, for a 10$^{7}$ M$_\odot$ black hole we expect the Keplerian
orbital period to be $\sim$10$^4$ s at a radius of 10~$\mathrm{r_g}$. Thus,
selecting 2.5 ks as the excess maps time resolution, enables
us to completely oversample this typical time-scale.
This choice of time resolution, optimized for the 2001b
observation, was extended to the 2000 and 2001a data.

\begin{table}
\caption{
Spectral features of interest in the 4--9 keV band with
the selected band-passes and mean intensity.}
\label{tab:bandpasses}     
\centering          
\begin{tabular}{c c c }
\hline\hline             
 Feature & Energy band & $\langle I \rangle$ \\
         &  (keV)      & ($10^{-5}\, \mathrm{ph \, s^{-1} cm^{-2}}$)\\
\hline                       
 red        & 5.3--5.4 & 0.6   \\
 wing       & 5.8--6.1 & 2    \\
 core (K$\alpha$)& 6.2--6.5 & 5.3\\
 K$\beta$   & 6.8--7.1 & 1.2   \\   
 Red+Wing   & 5.3--6.1 & 3.2   \\   
\hline                        
\end{tabular}
\end{table}

   \begin{figure*}
   \centering
    \includegraphics[height=10cm,width=3cm]{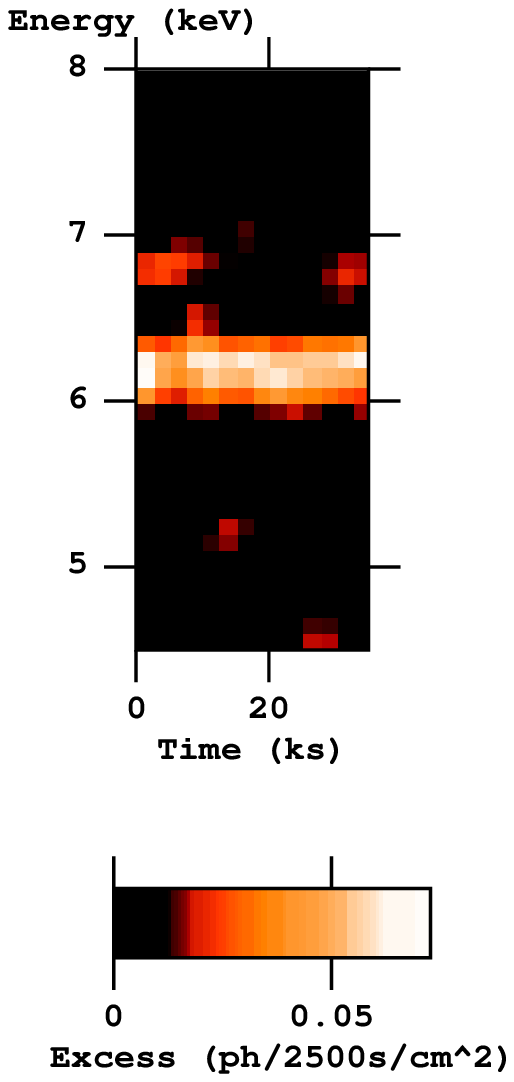}
    \includegraphics[height=10cm,width=12cm]{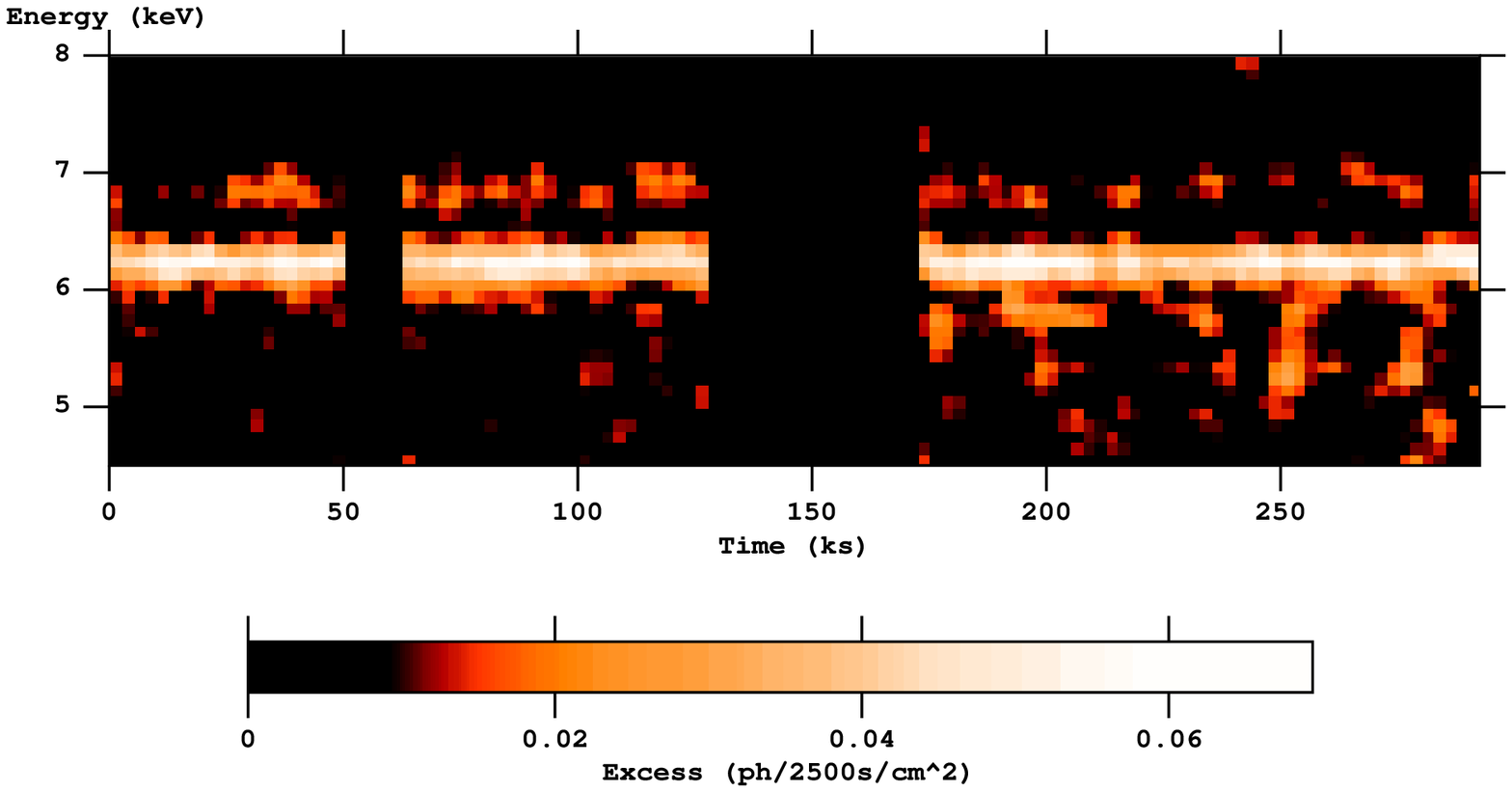}
   \caption{
The excess emission maps of the 4--9 keV band in the time-energy plane at 2.5 ks time resolution. The images have
been smoothed. Since the 6.4 keV line core is very strong and stable, the color map is adjusted to saturate the line core and
allow lower surface brightness features to be visible. {\it Left panel}: excess emission map from the 2000 observation. 
{\it Right panel}: excess emission map from the 2001a and 2001b observations.
}
              \label{f3}
    \end{figure*}

\section{Excess emission maps}

Energy spectra for a duration of the chosen exposure time
(2.5 ks) are extracted in time sequence. 14 spectra are obtained
from the 2000 observation, 46 and 48 spectra are
obtained from 2001a and 2001b observations respectively.
For each spectrum the continuum is determined and subtracted.
The residuals in counts unit are corrected for the
detector response and put together in time sequence to
construct an image in the time-energy plane.

\subsection{Continuum subtraction}

The continuum model is assumed to be always a simple absorbed 
power-law, throughout all the observations. For each spectrum the energy band 
of the observed spectral features (i.e. 5--7 keV) is excluded during the 
continuum fit. The 4--5 keV and 7--9 keV data 
are rebinned so that each channel contains more than 50 counts to enable 
the use of the $\chi^2$ minimization process when performing 
spectral fitting and to ensure that the high energy end of the data 
(7--9 keV) have enough statistical weight.
Because of the chosen low energy bound (4 keV), the fit
is not sensitive to cold absorption. Thus the cold absorption
column density is fixed to the time-averaged spectrum
value ($N_{\rm H} \simeq 2.5 \times 10^{22}$ cm$^{-2}$). 
Each 4--9 keV spectrum at 100 eV energy resolution is then 
fitted with its best-fit continuum model and residuals are used to 
construct the excess emission map in the time-energy plane.

Once all the continuum spectral fits have been done, we checked 
if continuum changes could affect our measurements of the line fluxes.
The mean power-law slopes during the three observations are 
1.85, 1.75 and 1.79 respectively, with standard deviations of 
0.08, 0.1 and 0.09. The power-law slopes are indeed quite constant, consistent 
with values obtained from the mean spectrum (\S 3.1), which result in a very 
marginal effect ($<0.1$\%) in the flux measurement of the narrow 
features we found here. 
These power-law continuum slopes are also in agreement with those previously 
found in observations using other instruments with overlapping spectral 
coverage, like \emph{BeppoSAX} (De Rosa et al. \cite{derosa}), 
\emph{ASCA}, \emph{RXTE} and \emph{Chandra} (Kaspi et al. \cite{kaspi}).

\subsection{Image smoothing}

As discussed in Iwasawa, Miniutti \& Fabian (\cite{iwasawa2004}), 
if the data are acquired
continuously and the characteristic time-scale of
any variation in a feature of interest is longer than the
sampling time (i.e. the time resolution), it is possible to
suppress random noise between neighboring pixels by applying
a low-pass filter. A circular Gaussian filter is used
with $\sigma$=0.85 pixel (200 eV in energy and 5 ks in time,
FWHM). The excess map Gaussian-filtered images for each
observation are shown in Fig.~\ref{f3}. Systematic variations are
observed in the 5.3--5.4 keV and in the 5.8--6.1 keV energy
bands of the 2001b observation. However, the image
filtering can slightly smear these narrow features and reduce
their intensity.

\section{Results}

\subsection{Light curves of the individual spectral features}

\begin{figure}
   \centering
    \includegraphics[width=7cm,height=8.5cm,angle=270]{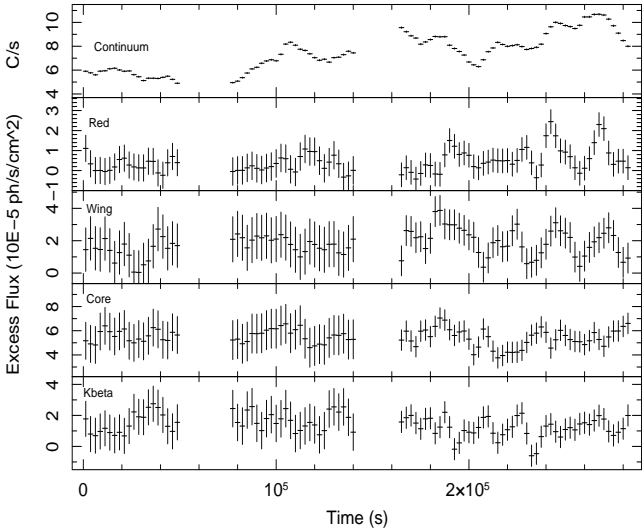}
   \caption{
The light curves of the total 0.3--10 keV continuum
flux and of the four spectral features (Tab.~2) extracted from
the excess maps of the 2001a and 2001b observations (Fig.~\ref{f3},
Right panel), with errors computed from simulations. The time
resolution is 2.5 ks.
}
              \label{f4}
    \end{figure}

Light curves of the four emission features are extracted
from the excess map filtered images. The selected band-passes
are listed in Tab.~2. During the image filtering process
individual pixels lose their independence to the neighbouring
ones. This means that a simple counting statistics
may be inappropriate for estimating the features light
curves errors. For this reason the estimation of the errors
has been done by extensive Monte Carlo simulations. We
implemented 1000 simulations following the same procedure
in making the excess map images. In the simulations
all the spectral features parameters and the power-law
slope are assumed to be constant, while letting the power-law
normalization vary according to the 0.3--10 keV light
curve. Light curves of individual spectral features have
been extracted from each simulation and their mean values
and variances recorded. The square root of the mean
of the variances (i.e. the dispersion) has been regarded as
the light curves error. In Fig.~\ref{f4} the emission features
light curves for the 2001a and 2001b observations are shown.
The 2000 observation light curves are not reported because
they do not show any sign of variability. The most
intense variations are registered in the light curves of the
``red" (E=5.3--5.4 keV) and ``wing" (E=5.8--6.1 keV) features
during the 2001b observation. The observed peaks
seem to follow the same kind of variability pattern and, as
shown in more details below, appear to be in phase with
the continuum emission.
In order to check the significance of the observed variability
we extracted both real and simulated data light curves
in the entire 5.3--6.1 keV band, i.e. of the ``red+wing"
structure. Then we compared the $\chi^2$ values against a constant 
hypothesis for the real data and the 1000 simulations; equivalent 
results can be derived comparing the variances directly. Only 73 of
the simulations show variability at the same level or greater than 
the real data, therefore we get a variability confidence level of 93\%.\\  
The light curves of the excess emission features in the
\mbox{5--6} keV energy band (red, wing and red+wing) seem to
show a variability pattern with a recurrence of the flux
peaks on time-scales of 27 ks. We further investigated how
it is likely to occur by chance applying a method that makes
use of the 1000 Monte Carlo simulations. We folded the
real data light curve with the interval of 27 ks and we fitted
it with a constant. We obtained a $\chi_{r}^{2}$ = 88 for 19 degrees
of freedom. We did the same to the simulated red+wing
light curves but, this time, folding in $n = 9$ trial periods,
from 17 to 37 ks at intervals of 2.5 ks, and recorded the
$\chi_{i}^{2}$ values. If $N$ is the total number of simulated red+wing
light curves for which $\chi_{i}^{2} \geq \chi_{r}^{2}$, the confidence 
level can be derived as $(1-\frac{N}{1000 \cdot n})$. Only $N$ = 54 
of the simulated red+wing light curves folded at the trial periods 
show chi-square values greater than the real one. Therefore, we
could derive a confidence level for the recurrence pattern
on the 27 ks time scale of 99.4\%.

Finally, we checked that the continuum above 7 keV (which carries the photons 
eventually responsible for the Fe line production) varies following the same
pattern  of the 0.3--10 keV band (Fig.~\ref{f4}, Top panel).

\subsection{Correlation with the continuum light curve}

In the 0.3--10 keV light curve of observation 2001b
(Fig.~\ref{f5}, Upper panel) flux variations of $\sim$30\%
are visible with four peaks separated by approximately equal time intervals.
Given such peculiar time series shape, we focused on
this observation and searched for some typical time-scale in
the variability pattern. Thus, we applied the \emph{efsearch} task 
(in \emph{Xronos}), which  searches for periodicities in a time series 
calculating the maximum chi-square of the folded light curve over a range of 
periods. We found a typical time-scale for variability of 26.6$\pm$2.2 ks. We 
then removed the underlying long-term 
variability trend by subtracting a 4th degree polynomial to the 0.3--10
keV continuum light curve (see Fig.~\ref{f5}, Middle
panel). The polynomial has been determined using the \emph{lcurve} task 
(in \emph{Xronos}), which makes use of the least-square technique. Applying
again the \emph{efsearch} task we found a typical time-scale for
short-term variability of $\sim$27.4 ks\footnote{It should be noted that the 
variability PSD study of this \emph{XMM-Newton} dataset by Markowitz 
(\cite{markowitz}) suggests an excess of power around 
4$\times$10$^{-5}$ Hz (corresponding to 
about 25 ks) during the 2001b observation (square symbols in his Fig.~3).}.
The peaks observed in the continuum light curve seem
to appear at the same times at which those observed in
the ``wing" and ``red" light curves do. In order to look
for some correlation between the continuum and the 5.3--6.1 keV 
(``red+wing") feature flux we computed the cross
correlation function (CCF) between the two time series,
where the input continuum light curve is the ``de-trended"
one. It is reported in Fig.~\ref{f6} as a function of time delay,
measured with respect to the continuum flux variations.
No delay is evident, with an estimated error at the peak of
2.5 ks. The continuum and ``red+wing" fluxes seem to
show a correlation, with peak value  0.7. To estimate the
significance of the correlation we computed the CCFs between
the continuum and the simulated ``red+wing" light
curves. If $N$ is the number of simulated light curves which
have a higher cross correlation than the real one, the significance 
of the correlation is ($1-N/1000$). Applying this
method we found a confidence level greater than 99.9\%.

\subsection{High/Low flux state line profiles}

\begin{figure}
   \centering
    \includegraphics[width=7cm,height=8cm,angle=270]{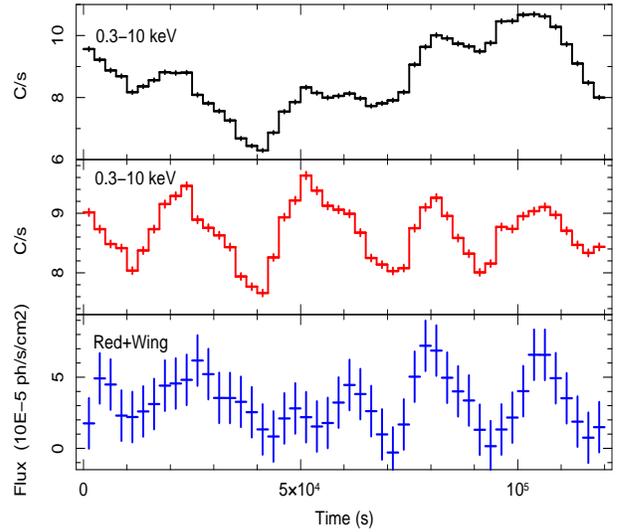}
   \caption{
\emph{Upper panel}: The 0.3--10 keV light curve of NGC~3783
during 2001b observation at 2.5 ks time resolution. \emph{Middle
panel}: The 0.3--10 keV light curve of NGC~3783 during the 2001b
observation after subtraction of a 4th degree polynomial (long-term
variations) at 2.5 ks time resolution. \emph{Lower panel}: The
5.3--6.1 keV (``red+wing" energy band) light curve extracted
from the excess emission map of the 2001b observation (Fig.~\ref{f3}, Right panel) at 2.5 ks time resolution.}
              \label{f5}
    \end{figure}

\begin{figure}
   \centering

    \includegraphics[width=5cm,height=6cm,angle=270]{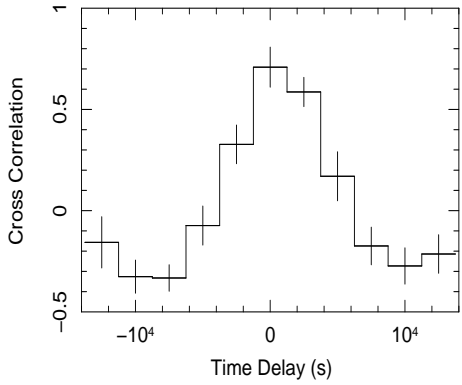}
   
   \caption{
The cross correlation function calculated between the
de-trended 0.3--10 keV continuum light curve and the 5.3--6.1
keV feature (``red+wing") light curve.}
  \label{f6}
    \end{figure}

\begin{figure}
   \centering
    \includegraphics[width=6cm,height=8cm,angle=270]{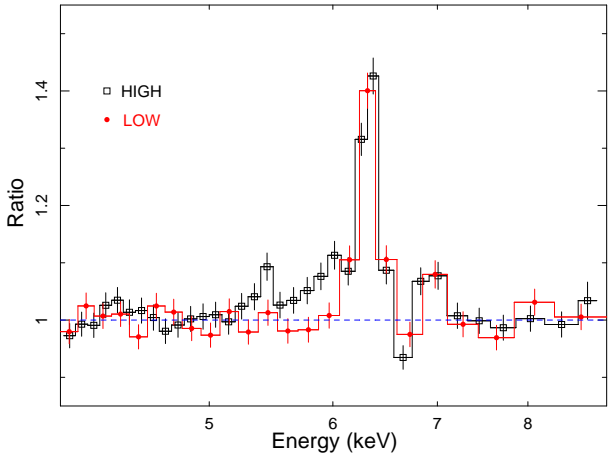}
   \caption{
The Fe~K line profile during the High flux (open squares)
and the Low flux (solid circles) phases of the 5.3--6.1 keV feature.
The ratios are computed against the best-fitting continuum
model. The energies are in the observer frame.
}
              \label{f7}
    \end{figure}

Looking at the ``red+wing'' light curve (Fig.~\ref{f5}, Lower
panel) we constructed two spectra from the integrated high and
low flux intervals to verify the variability in this energy
band. The line profiles are shown in Fig.~\ref{f7} where the ratio
between the data and a simple power-law plus cold absorption
model is shown. While the 6.4 keV and 6.9 keV
features remain the same, a small increase of counts is visible
in the 5.3--5.4 keV and \mbox{5.8--6.1} keV bands in the
high flux state. Adding an emission Gaussian model to the
simple power law plus Gaussian line (at the Fe~K$\alpha$ energy)
model in the high flux state integrated spectra improves
the $\chi^2$ of 18. Thus the significance of the excess is 
$\sim$99.9\%.

\section{Discussion}

In the 2001b observation, NGC~3783 clearly exhibits continuum emission
with two different time-scales: a long-term modulation with variations
up to a factor 2 on intervals greater than 60 ks is superimposed to a
shorter one on a time-scale of 27 ks, with modulations of 30\% of the
average value.
According to an estimated black hole mass of $\mathrm{M_{BH}}=(3.0 \pm 0.5) 
\times 10^7~\mathrm{M_\odot}$ (Peterson et al. \cite{peterson}), the latter 
modulation 
occurs with a characteristic time-scale corresponding to the orbital period at 
\mbox{$\sim$9--10~$\mathrm{r_g}$} (e.g. Bardeen, Press \& Teukolsky 
(\cite{bardeen})). 
It should be noted, however, that a previous mass estimate by Onken \& Peterson
(\cite{onken}) gave the value of 
$\mathrm{M_{BH}}= (8.7 \pm 1.1)\times 10^6~\mathrm{M_\odot}$,
that would correspond to $\sim$20~$\mathrm{r_g}$. 
These mass discrepancies are mainly due to a different scaling of the virial 
relation in performing reverberation mapping measurements. However, the former 
estimate is more accurate because it has been calibrated to the 
$\mathrm{M_{BH}}-\sigma$ relation and thus we will adopt this value in the 
following discussion. The power spectral density of the source is
consistent with a red-noise shape (Markowitz \cite{markowitz}) where 
variability mainly occurs on intervals of the order of days. 
Here we identify an additional (additive) shorter time-scale component 
(see the footnote 1), most likely produced within the 
innermost accretion flow/corona system.

The most remarkable result of our analysis is the detection of
redshifted (5.3--6.1~keV) Fe K emission and of its variability. 
The redshifted emission appears to respond only to the shorter 
$\sim$27~ks time-scale modulation and shows a good correlation with the
continuum with a time-lag consistent with zero within the errors
($\Delta\tau\sim 1.25$~ks). This indicates 
that the continuum modulation on this time interval is likely to induce 
Fe K emission from dense material close to the black hole (which explains 
the observed redshift of the emission feature); moreover the lack 
of time-lags implies that the distance between the sites of continuum 
and line production is smaller than c$\Delta\tau \sim 4\times 
10^{13}$~cm~$\sim 8~\mathrm{r_g}$ (for the black hole mass given above). 
We are therefore most likely observing emission from the innermost 
accretion flow in both the continuum and line emission (corona and disc).

As discussed above, the variability time-scale suggests we are looking
at emission from around $\sim$9--10~$\mathrm{r_g}$. As a consistency check, we
fitted the time-averaged spectrum by including a \emph{diskline}
component to account for the redshifted features. We forced the
emission region to be an annulus of $\Delta$r$= 0.5~\mathrm{r_g}$ with uniform
emissivity because the purpose of this test is to assess the
approximate location of the line-emitting region. We obtain good fits
with an almost face-on disc ($i = 11\pm 4^{\circ}$) and an annulus at 
9--15~$\mathrm{r_g}$  depending on the assumed Fe line rest-frame 
energy (from neutral 
at 6.4~keV to highly ionized at 6.97~keV). For all cases we tested, the 
statistical improvement is of $\Delta\chi^2 \sim 16$ for 3 additional degrees 
of freedom, corresponding to a confidence level of 99.7\%. 
This fit shows that the redshifted Fe line emission we detected is indeed 
consistent in shape with being produced around $10~\mathrm{r_g}$, 
where the disc 
orbital period is of the order of 27~ks, which agrees well with the 
correlated (and zero-lag) variability of the two components.

The interpretation of our results is however not straightforward. The
quasi-sinusoidal modulations of the continuum and line emission (see
Fig.~\ref{f5}) would suggest the presence of a localized co-rotating flare
above the accretion disc which irradiates a small spot on its surface. 
The intensity modulations we see (Fig.~\ref{f5}) would then
be produced by Doppler beaming effects (acting on both the flare and
spot emission, i.e. on both continuum and line) and the characteristic
time-scale of 27~ks would be identified with the orbital period (because of
gravitational time dilation, the period measured by an observer on the disc at 
$10~\mathrm{r_g}$ would be shorter by $\sim$10\%). As
demonstrated above, a flare/spot system orbiting the black hole at
$\sim$10~$\mathrm{r_g}$ would also produce a time-averaged line profile in
agreement with the observations. However, such a model makes definite
prediction on the Fe line energy modulation within one orbital
period. In this framework, the orbiting spot on the accretion disc
would also give rise to energy modulations of the Fe line due to the
Doppler effect and such energy modulation is barely seen in the data (see
Fig.~\ref{f3}). We stress that the adopted time resolution (2.5~ks) is good
enough to detect energy modulations with a characteristic 27~ks
time-scale. This has been demonstrated with \emph{XMM-Newton} in the
case of NGC~3516, where the modulation occurs on a very similar
time-scale (Iwasawa et al. \cite{iwasawa2004}; 
see also Dov{\v c}iak et al. \cite{dovciak} for
theoretical models). Therefore, the lack of Fe line energy modulation
disfavours the orbiting flare/spot interpretation for NGC~3783.

However, a variability time-scale of the order of the orbital one at 
a given radius does not necessarily imply the motion of a point-like
X-ray source. In fact, since the orbital time-scale is the fastest at
a given disc radius, and since the observed time-scale of $\sim$27~ks
corresponds to the orbital period at $\sim$10~$\mathrm{r_g}$, one 
could argue that the data only imply that the X-ray variability likely 
originates from within $\sim$10~$\mathrm{r_g}$. The apparent 
recurrence in the X-ray continuum modulation may not necessarily be related to
a real physical periodicity, especially considering the limited length of the
observation (only four putative cycles are detected).

A possible explanation for the observed behaviour is that the X-ray
continuum source(s) (located within $\sim$10~$\mathrm{r_g}$ from the 
center) irradiates the whole accretion disc, but only a ring-like structure
around $\mathrm{10~r_g}$ is responsible for the fluorescent Fe emission. This
is possible if the bulk of the accretion disc is so highly ionized
that little Fe line is produced, while an over-dense (and therefore
lower ionization) structure is responsible for the fluorescent
emission. Such an over-dense region could have an approximate
ring-like geometry if it is for instance associated with a spiral-wave
density perturbation. In this case, the Fe emitting region is extended
in the azimuthal direction and we do not expect strong energy
modulation with time, whatever the origin of the continuum
modulation. We point out that spiral density distributions could
result from the ordered magnetic fields in the inner region of the
disc and that the energy dissipation (via e.g. magnetic reconnection)
could be enhanced there, thereby providing a common site for the
production of the X-ray continuum and the Fe line (e.g. Machida \&
Matsumoto \cite{machida}).

On the other hand, if the apparent recurrence is in fact real, it
is worth noting that the continuous theoretical effort in
understanding the origin of quasi-periodic oscillations (QPO) in
neutron star and black hole systems provides a wealth of mechanisms
inducing quasi-periodic variability, although none is firmly
established (Psaltis \cite{psaltis}; Kato \cite{kato}; Rezzolla et al. 
\cite{rezzolla}; Lee et al. \cite{lee}; Zycki \& Sobolewska \cite{zycki} 
and many others). A
connection between QPO phase and Fe line intensity has been previously
claimed in the Galactic black hole GRS~1915+105 (Miller \& Homan
\cite{miller2005}). Although in our case, the presence 
of a QPO cannot be claimed
because of the very small number of detected cycles, the analogy
is suggestive. In the case of GRS~1915+105, Miller \& Homan consider
that a warp in the inner disc, possibly due to Lense-Thirring
precession, may produce the observed QPO-Fe line connection
(e.g. Markovic \& Lamb \cite{markovic}). However, to produce the observed 
$\sim$15\% rms variability in the X-ray lightcurve, 
the black hole spin axis should be inclined with respect to the line of sight 
by at least $60^\circ$ (which is at odds with our inclination estimate
of $11\pm 4^\circ$ and with the Seyfert~1 nature of NGC~3783), and 
the tilt precession angle should be larger than 20$^\circ$--30$^\circ$ 
(Schnittman, Homan, Miller \cite{schnittman}). 
Both requirements make it highly unlikely 
that Lense-Thirring precession can successfully account for the observed 
modulations in NGC~3783.

While the origin of the coherent intensity modulation still remains
unclear, the correlated variation of the continuum and line emission
and the Fe line shape are consistent with an emission site at 
$\sim$10~$\mathrm{r_g}$.  Moreover, the fact that the iron line 
variability responds to the 27~ks time-scale modulation only implies that 
this short time-scale variation is somehow detached from the long-term
variability. The latter may be associated with perturbations in the
accretion disc propagating inwards from outer radii and modulating the
X-ray emitting region (Lyubarskii \cite{lyubarskii}), while the former seems to
genuinely originate in the inner disc.

Further observational data may help to clarify the complex phenomena 
related to the relativistic Fe line temporal evolution in Seyfert 1 galaxies. 
Our work makes it clear that higher quality data in the Fe band will be able 
to probe the innermost regions of accretion flows with high accuracy. 
Next generation of large collecting area X-ray missions such as 
\emph{XEUS} and \emph{Constellation-X} or even very long observations with 
\emph{XMM-Newton} will be crucial to fully exploit such potential.

\begin{acknowledgements}

This paper is based on observations obtained with the \emph{XMM-Newton} 
satellite, an ESA funded mission with contributions by ESA Member States and 
USA. We thank A. M\"uller, K. Nandra, L. Nicastro, P. O'Neill and M. Orlandini 
for useful discussions. MC, MD and GP acknowledge financial support from ASI 
under contract ASI/INAF I/023/05/0. The authors thank the anonymous referee for
suggestions that led to improvements in the paper.

\end{acknowledgements}


\begin{thebibliography}{}

\bibitem[1972]{bardeen} 
  Bardeen, J.~M., Press, W.~H., \& Teukolsky, S.~A.\ 1972, \apj, 178, 347 

\bibitem[2006]{brenneman} 
  Brenneman, L.~W., \& Reynolds, C.~S.\ 2006, ArXiv Astrophysics e-prints, 
arXiv:astro-ph/0608502

\bibitem[2002]{derosa} 
  De Rosa, A., Piro, L., Fiore, F. et al. \ 2002, \aap, 387, 838

\bibitem[2006]{demarco}
  De Marco, B., Cappi, M., Dadina, M., \& Palumbo, G.G.C., 2006, Astron. Nach., astro-ph/0610882 

\bibitem[2004]{dovciak} 
  Dov{\v c}iak, M., Bianchi, S., Guainazzi, M., Karas, V., \& Matt, G.\ 2004, \mnras, 350, 745 

\bibitem[2000]{fabian} 
  Fabian, A.~C., Iwasawa, K., Reynolds, C.~S., \& Young, A.~J.\ 2000, \pasp, 112, 1145 

\bibitem[2006]{guainazzi}
  Guainazzi, M., Bianchi, S., \& Dovciak, M.\ 2006, ArXiv Astrophysics e-prints, arXiv:astro-ph/0610151 

\bibitem[1996]{iwasawa1996} 
  Iwasawa, K., et al.\ 1996, \mnras, 282, 1038

\bibitem[2004]{iwasawa2004} 
  Iwasawa, K., Miniutti, G., \& Fabian, A.~C.\ 2004, \mnras, 355, 1073 

\bibitem[2001]{kaspi} 
  Kaspi, S., et al.\ 2001, \apj, 554, 216 

\bibitem[2001]{kato} 
  Kato, S.\ 2001, \pasj, 53, L37 

\bibitem[2004]{lee} 
  Lee W.H., Abramowicz M.A., \& Kluzniak W., 2004, ApJ, 603, L93

\bibitem[1997]{lyubarskii} 
  Lyubarskii, Y.~E.\ 1997, \mnras, 292, 679 

\bibitem[2003]{machida} 
  Machida, M., \& Matsumoto, R.\ 2003, \apj, 585, 429 

\bibitem[1998]{markovic} 
  Markovic D. \& Lamb F.K., 1998, ApJ, 507, 316

\bibitem[2005]{markowitz} 
  Markowitz, A.\ 2005, \apj, 635, 180 

\bibitem[2005]{miller2005} 
  Miller J.M. \& Homan J., 2005, ApJ, 618, L107

\bibitem[2006]{miller2006} 
  Miller, L., Turner, T.~J., Reeves, J.~N. et al. \ 2006, \aap, 453, L13 

\bibitem[2004]{miniutti2004} 
  Miniutti, G., \& Fabian, A.~C.\ 2004, \mnras, 349, 1435 

\bibitem[2006]{miniutti2006}
  Miniutti, G., et al.\ 2006, ArXiv Astrophysics e-prints, arXiv:astro-ph/0609521 

\bibitem[1997]{nandra1997}
  Nandra, K., George, I.~M., Mushotzky, R.~F., Turner, T.~J., \& Yaqoob, T.\ 1997, \apj, 477, 602 

\bibitem[2006]{nandra2006} 
  Nandra, K., O'Neill, P.~M., George, I.~M., Reeves, J.~N., \& Turner, T.~J.\ 2006, ArXiv Astrophysics e-prints, arXiv:astro-ph/0610585 

\bibitem[2006] {oneill}
  O'Neill, P. \& Nandra, K., 2006, in preparation

\bibitem[2002]{onken}
  Onken, C.~A., \& Peterson, B.~M.\ 2002, \apj, 572, 746 

\bibitem[2004]{peterson}
  Peterson, B.~M., et al.\ 2004, \apj, 613, 682 

\bibitem[2004]{ponti}
Ponti, G., Cappi, M., Dadina, M., \& Malaguti, G.\ 2004, \aap, 417, 451 

\bibitem[2001]{psaltis}
  Psaltis D., 2001, Adv Space Res., 28, 481

\bibitem[2004]{reeves}
  Reeves, J.~N., Nandra, K., George, I.~M. et al. \ 2004, \apj, 
602, 648

\bibitem[2003]{reynolds}
  Reynolds, C.~S., \& Nowak, M.~A.\ 2003, \physrep, 377, 389 

\bibitem[2003]{rezzolla}
  Rezzolla L., Yoshida S., Maccarone T.J., Zanotti O., 2003, MNRAS, 344, L37

\bibitem[2006]{schnittman}
  Schnittman J.D., Homan J., Miller J.M., 2006, ApJ, 642, 420

\bibitem[2006]{turner}
  Turner, T.~J., Miller, L., George, I.~M., \& Reeves, J.~N.\ 2006, \aap, 445, 59 

\bibitem[2001]{vaughan}
  Vaughan, S., \& Edelson, R.\ 2001, \apj, 548, 694 

\bibitem[2005]{zycki}
  Zycki P.T. \& Sobolewska M.A., 2005, MNRAS, 364, 891

\end{thebibliography}
\end{document}